\documentclass[prl,aps,twocolumn,superscriptaddress,times]{revtex4-1}
\usepackage{amssymb,amsfonts,amsmath,amsfonts,graphicx,mathptm, MnSymbol}
\usepackage[usenames,dvipsnames]{color}

{\catcode`\|=\active
  \gdef\Braket#1{\begingroup
\mathcode`\|32768\let|\BraVert\left<{#1}\right>\endgroup}}
\def\BraVert{\egroup\,\mid\,\bgroup}

\begin{document}

\title{Self-Stabilizing Measurements for Noisy Metrology}

\author{Sai Vinjanampathy}
\email{sai@quantumlah.org}
\affiliation{Centre for Quantum Technologies, National University of Singapore, 3 Science Drive 2, Singapore 117543.}

\begin{abstract} 
We present a protocol to perform self-stabilizing measurements on noisy qubits. We employ rapid purification in a rotating frame whose frequency is estimated and periodically updated via a Bayesian estimation scheme. The Bayesian estimation protocol employs the continuous measurement record to improve the estimate, which in turn purifies the qubit more. This procedure stabilizes the qubit. Such an adaptive measurement scheme serves the purpose of purifying the state, while minimally interfering with the phase estimation.
\end{abstract}
\maketitle
\section{Introduction}
Parameter estimation in noise environments is the key challenge for the practical realization of quantum metrology \cite{giovannetti2011advances,PhysRevLett.96.010401,stockton2004robust,demkowicz2009quantum,modi2011quantum,dowling2008quantum}. Depending on the choice of input states and Hamiltonians, quantum metrological schemes have demonstrated an advantage over classical metrological schemes \cite{dowling2008quantum}. The figure of merit to judge the goodness of a metrological scheme is the variance of the estimated phase, denoted by $\Delta^{2}\varphi_{\rm{est}}$. One expects that larger the number of particles $N$ that are involved in acquiring the unknown phase, the smaller the phase variance should be. Furthermore, increasing the number of times $\nu$ that the measurement is repeated is also expected to decrease the estimated phase variance. This intuition is seen to be true, for instance from the quantum Cram{\'e}r-Rao bound \cite{braunstein1994statistical,luo2000quantum,paris2009quantum} that states that the variance of unbiassed estimators of a parameter $\varphi$ scales as 
\begin{gather}
\Delta^2\varphi_{\rm{est}}\geq\frac{1}{\nu \mathcal{F}_{Q}}.
\end{gather}
Here, $\mathcal{F}_Q$ stands for the quantum Fisher information (QFI) and is related to $N$. For instance, if coherent light $\vert\alpha\rangle$ is used in an interferometric scheme with the Hamiltonian $\varphi\hat{n}$ being linear in the number operator $\hat{n}$, $\mathcal{F}_Q$ can be as big as $\vert\alpha\vert^2$, the average number of photons in the coherent state. On the other hand, quadratic scaling of QFI for linear Hamiltonian interactions has been studied theoretically \cite{anisimov2010quantum,demkowicz2009quantum,demkowicz2012elusive} and experimentally \cite{napolitano2011interaction} (also see \cite{demkowicz2012elusive}). Two problems remain in the way of the implementation of realistic quantum metrology in the presence of noise. The first is a theoretical challenge concerning asymptotic theoretical bounds and the second challenge involves quantum metrology in the presence of decohering environments. An asymptotic bound on the estimated phase variance is placed by the quantum Cram{\'e}r-Rao bound noted above. There has been some work to address realistic bounds in for finite number of measurements \cite{tsang2012ziv,gao2012generalized}. If the parameter is acquired by the action of a Hamiltonian $\mathrm{G}$, the quantum Fisher information defined as
\begin{gather}\label{hubnereq}
\mathcal{F}_{Q}=\sum_{j,k} \frac{(\lambda_j-\lambda_k)^2}{\lambda_j+\lambda_k}\vert\langle\Psi_j\vert \mathrm{G}\vert\Psi_k\rangle\vert^2.
\end{gather}
This formula \cite{luo2000quantum} involves the instantaneous eigenstates of the density operator $\{\vert\Psi_j\rangle\}$ and the corresponding eigenvaues $\{\lambda_j\}$. Since we wish to decrease $\Delta\varphi_{est}$, the asymptotic analysis indicates that increasing Fisher information will reduce $\Delta\varphi$.To this end, we note three properties of $\mathcal{F}_Q$: firstly since $\mathcal{F}_Q\propto \mathrm{G}$, to increase quantum Fisher information, it suffices to increase the Hamiltonian strength. Let this strength be represented by the norm of the Hamiltonian $\Vert\mathrm{G}\Vert$. But this involves increasing the energy needed to implement the Hamiltonian, which is undesirable. We hence set $\Vert\mathrm{G}\Vert=1$, corresponding to investigating schemes that have the same (fixed) Hamiltonian strength. Secondly, we note that since $\mathcal{F}_Q$ is convex, pure state probes are always better than mixed state probes. Thirdly, we note that for a single qubit state $\varrho=(\mathbb{I}+r.\sigma)/2$, Eq.(\ref{hubnereq}) suggests that the Hamiltonian that maximizes the quantum Fisher information is perpendicular to $r$, namely $\mathrm{G}=r_{\perp}.\sigma$. Since the Hamiltonian is in general not controllable, we will, without loss of generalization, take the Hamiltonian to be $\sigma_x$. Motivated by these three criteria, in this paper, we will propose a method to perform self-stabilizing phase measurements on a decohering qubit.

The scheme consists of actively purifying the qubit as it acquires an unknown phase. We employ continuous measurement feedback control \cite{,jacobs2006straightforward,wiseman2010quantum} to implement the scheme. Though we are motivated by the formula for quantum Fisher information in Eq.(\ref{hubnereq}) to derive the three desired criterion for a ``good" feedback controlled metrological scheme, we will not employ quantum Fisher information as the figure of merit. This is due to the fact that quantum Fisher information is an asymptotic bound. The continuous measurements used to implement the control scheme are represented in terms of a stochastic master equation, namely, 
\begin{gather}\label{SME}
d\varrho=-\frac{i}{\hbar}[\varphi\mathrm{G},\varrho]dt+\displaystyle\sum_{j=1}^{3}\frac{\gamma_j}{2}\mathcal{D}[\sigma_j]\rho dt+\mathcal{D}[c]\rho dt+\sqrt{\eta}\mathcal{H}[c]\rho d\mathbb{W}.
\end{gather}
While the first term in the equation above represents the Hamiltonian evolution with the unknown phase, the next term represents decoherence of the qubit represented by the action of three Pauli operators with damping factors $\gamma_j$. Here, $\mathcal{D}[c]\varrho:=c\varrho c^{\dagger}-(c^{\dagger}c\rho+\rho c^{\dagger}c)/2$ represents the Lindblad super operator corresponding to decoherence or measurement back-action and $\mathcal{H}[c]\varrho :=c\varrho+\varrho c^{\dagger}-\langle c+c^{\dagger}\rangle\rho$ corresponds to the information gain due to the measurement. $\eta$ represents the measurement's detector efficiency, $\eta=1$ representing a unit efficiency detection process. $d\mathbb{W}$ is a Wiener increment \cite{jacobs2010stochastic,gardiner1985handbook} given by zero mean and $\llangle d\mathbb{W}^2\rrangle=dt$. The measurement record for this process can be written as
\begin{gather}
dy(t) = \frac{\langle c+c^{\dagger}\rangle}{2}dt+\frac{d\mathbb{W}}{\sqrt{4\eta}}.
\end{gather}
Such a continuous measurement and control of a quantum system has been studied and demonstrated in a variety of physical systems including quantum dots \cite{korotkov1999continuous}, nano mechanics \cite{truitt2007efficient,suh2010parametric,jacobs2011real}, circuit quantum electrodynamics (CQED) \cite{blais2004cavity,siddiqi2004rf,vijay2012stabilizing,friedman2000quantum} and cavity quantum electrodynamics \cite{brune1992manipulation}. 
\section{Rapid Purification and Self Stabilization}
Our protocol will involve purifying a qubit that is decohering as it gathers information about the unknown phase by improving its purity. If we were interested \textit{not} in phase estimation, but in simply purification, efficient algorithms to purify using continuous measurement quantum control already exist. In particular, several authors have investigated the purification speed arising from \textit{rapid purification} protocols \cite{combes2006rapid,combes2008rapid,wiseman2006reconsidering,combes2010replacing}. Such protocols aim to purify a qubit using Hamiltonian feedback and continuous measurements as quickly as possible. The Jacobs protocol involves an adaptive measurement scheme so that the measurement is always perpendicular to the state. The evolution of the linear entropy $S_L=1-\mathrm{tr}(\varrho^2)$ for the evolution of a qubit subject to an adaptive measurement is given by
\begin{gather}
dS_{L}=-2(r.dr_1dt+r.dr_2d\mathbb{W}).
\end{gather}
Here, we have written $d\varrho=dr_1.\sigma dt+dr_2.\sigma d\mathbb{dW}$ for brevity. From Eq.(\ref{SME}), it is clear that $dr_2$ depends entirely on the choice of the measurement operator. If this operator is chosen to be perpendicular to the instantaneous Bloch vector $r$, then the evolution of the linear entropy is deterministic and entirely dictated by $dr_1$. In the absence of decoherence (\textit{i.e.,} $\gamma_i=0$), for a measurement $c=\sqrt{\kappa/2}X$,the evolution of the linear entropy is given by \cite{combes2006rapid}
\begin{gather}
dS_L=-2\kappa\mathrm{tr}[X\varrho X\varrho]dt.
\end{gather}
This is solved to yield $S_L(t)=S_L(0)\exp(-2\kappa\mathrm{tr}[X\varrho X\varrho]t)$, purifying the qubit rapidly.

If the phase $\varphi$ were known, the rapid purification protocol could be implemented in a rotating basis. But, since $\varphi$ is the unknown phase we wish to measure, our scheme will involve implementing rapid purification in a rotating frame, whose frequency is the estimated phase $\varphi_{est}$. To estimate the phase, we will employ a Bayesian parameter inference from continuously monitored systems discussed in the next section. Every $m$ cycles, the measurement record is used to perform a Bayesian update and the updated estimator is used to calculate the average position of the density matrix for the next $m$ cycles of measurements. This allows us to perform rapid purification in the rotating frame of the estimated phase. At the beginning of the protocol, since the prior probability density is assumed to be flat, corresponding to the absence of any knowledge about the unknown phase $\varphi$. We hence choose a fixed axis (the axis we prepared the state in) and perform measurements perpendicular to that axis for the first $m$ cycles. Since we do not wish to interfere with the Hamiltonian, we will indeed pick an axis that is mutually perpendicular to the qubit state and the Hamiltonian at any given time. At the end of that block, we estimate the unknown phase by computing $\rm P[\varphi\vert y(t)]$, the conditional probability given the measurement record $y(t)$. If the corresponding variance $\Delta\varphi_{est}$ is less than a given tolerance $\epsilon$, another block of simulations and measurements are performed. An alternative stopping criterion for this protocol involves a predetermined total number of steps. This might be suitable if $\gamma_j$ are especially strong causing the qubit to eventually decohere completely. Note that though the first block of measurements has the effect of \textit{not} purifying the state in general, the data obtained via static measurements will aid in the implementation of a rapid purification scheme. With each block of evolution, the Bayesian estimate \cite{jaynes2003probability} of the unknown phase $\varphi_{set}$ will get closer to the true phase, causing the next block of simulated adaptive measurements to be closer to the ``ideal" adaptive measurement. This procedure hence has the effect of purifying the qubit \textbf{and} making the variance in the estimated phase smaller with each passing block. The steps of the protocol can be summarized as follows:
\begin{enumerate}
\item On the first block, simulate $m$ cycles of the state evolution with a static measurement operator, $\sigma_z$.
\item Estimate the phase at the end of the first block.
\item Use the estimated phase to compute the average trajectory for the next block. Use this to identify the measurement vectors that are mutually perpendicular to $\rm G$ and $\varrho$. 
\item Repeat previous step until $\Delta\varphi_{est}<\epsilon$.
\end{enumerate}
In the next section, we will take up the task of estimating the unknown phase.
\section{Bayesian Estimation from a Continuous Record}
The central task at the end of each block of evolution of the protocol outlined in the previous section is the estimation of an unknown phase $\varphi$, given a measurement record $y(t)$ \cite{geremia2003quantum,gammelmark2013bayesian}. This issue was studied in \cite{gammelmark2013bayesian} and is summarized in this section. Baye's law applied to the measurement record $y(t)$ states that 
\begin{gather}
\rm P[\varphi\vert y(t)]=\frac{\rm P[y(t)\vert\varphi]\rm P[\varphi]}{\rm P[y(t)]}.
\end{gather}
Here $\rm P[\varphi\vert y(t)]$ represents the conditional probability density for the parameter $\varphi$, given the data $y(t)$, $\rm P[\varphi]$ represents the prior probability distribution of the unknown parameter $\varphi$ and $\rm P[y(t)]=\int d\varphi\rm P[y(t)\vert\varphi]\rm P[\varphi]$. A log likelihood function $l(\varphi\vert y(t))=\log(L[\varphi\vert y(t)])$ can be defined in terms of the likelihood function, given by
\begin{gather}
L[\varphi\vert y(t)]=\frac{P[\varphi\vert y(t)]}{P_0[\varphi]}.
\end{gather}
Here $P_0[\varphi]$ is a convenient choice of normalization. While the probability distribution $P[y(t)\vert\varphi]$ informs us about the probability of generating a measurement record for a given parameter, likelihood functions inform us of the opposite: the likelihood of a parameter given a measurement record. To apply this to continuous measurements, we first note that the effect of the measurement operator outcome $x$ on the state can be written as
\begin{gather}
\rho\vert x=\frac{\Omega(x)\rho\Omega^{\dagger}(x)}{p(x)},
\end{gather}
where the probability of observing this outcome is given by $p_x=\mathrm{tr}[\Omega(x)\rho\Omega^{\dagger}(x)]$. The probability operators $\Omega^{\dagger}(x)\Omega(x)$ are normalized as
\begin{gather}
\displaystyle\int dx \Omega^{\dagger}(x)\Omega(x)=\mathbb{I}. 
\end{gather}
Furthermore, introducing an ``ostensible probability" $p_0(x)$, the authors in \cite{gammelmark2013bayesian} define a new set of POVMS, namely $\Omega(x)\rightarrow\Omega(x)/\sqrt{p_0(x)}$ so that the normalization condition above is modified to
\begin{gather}
\displaystyle\int dx p_0(x) \Omega^{\dagger}(x)\Omega(x)=\mathbb{I}. 
\end{gather}
This allows us to define a new set of states $\tilde{\rho}\vert x=\Omega(x)\rho\Omega^{\dagger}(x)$, whose trace now depends on $p_0(x)$. The role of $p_0(x)dx$ is to provide a reference measure on the set of measurement outcomes. Note that the trace of $\tilde{\varrho}\vert x$ now explicitly depends on the measurement record and does not change its dependance on $\varphi$ for various measurement outcomes. Hence, it was pointed out that it can serve as a good likelihood function. Hence, at a given time $t$, this analysis leads to the likelihood function being defined as $L(t)=\mathrm{tr}\{\tilde{\rho}(t)\}$. Here $\tilde{\rho}\vert x$ at the time $t$ is written as $\tilde{\rho}(t)$ for brevity. $L(t)$ obeys the evolution equation
\begin{gather}
dL(t)=\mathrm{tr}\{\mathcal{H}[c]\tilde{\rho}(t)\}dy(t)
\end{gather}
Returning to the protocol described in the previous section, a static (time-independant) measurement operator $c=\sigma_z$ is chosen for the first block of evolution. The continuous monitoring of the decohering qubit (assumed to decohere under thermal Lindbladians at a temperature $\beta^{-1}$) is simulated and the corresponding measurement record $dy(t)$ is employed to update the likelihood function. At the end of the first block of evolution, the unknown phase is estimated as 
\begin{gather}
\varphi_{est}=\displaystyle\int d\varphi \varphi \rm P[\varphi\vert y(t)],
\end{gather}
where $P[\varphi\vert y(t)]$, the probability density is given by
\begin{gather}
P[\varphi\vert y(t)]=\displaystyle\frac{L[y(t)\vert \varphi]P[\varphi]}{\int d\varphi L[y(t)\vert \varphi]P[\varphi]}.
\end{gather}
Furthermore, the variance is estimated directly from the probability density $P[\varphi\vert y(t)]$ as 
\begin{gather}
\Delta^2\varphi_{est}=\displaystyle\int d\varphi \{\varphi-\varphi_{est}\}^2 \rm P[\varphi\vert y(t)],
\end{gather}
Now, given $\varphi_{set}$, the next cycle of measurement directions is simulated to implement rapid purification over the next block of evolution. Since we cannot know the precise trajectory of the future blocks of evolution, we have to use the average equation to simulate the evolution of the Bloch vector. Since this will only approximately be the ``correct" feedback scheme (both due to $\Delta\varphi_{set}$ and due to the average equation), repeated cycles of estimation and evolution might be needed. We simulate such an evolution in the next section.
\section{Results and Discussion}
We consider a qubit undergoing evolution nuder a thermal Lindbladian, namely 
\begin{widetext}
\begin{gather}
d\varrho=-\frac{i}{\hbar}[\varphi\mathrm{G},\varrho]dt+\gamma\bar{n}\mathcal{D}[\sigma_+]\rho dt+\gamma(1+\bar{n})\mathcal{D}[\sigma_-]\rho dt+\mathcal{D}[c]\rho dt+\sqrt{\eta}\mathcal{H}[c]\rho d\mathbb{W}.
\end{gather}
\end{widetext}
Here $\bar{n}$ is the average number of thermal phonons in the qubit at thermal equilibrium. Fig.(1) shows the simulation of three cycles of evolution for realistic damping factors derived for a good CQED qubit. The damping corresponds to parameters given in \cite{vijay2012stabilizing}. The simulation reveals that for good control, where the measurement strength $\kappa$ was much larger than the qubit damping rates, the Bayesian estimator is able to estimate the unknown phase within a couple of cycles. In contrast, in the regime of bad control, when $\kappa$ and $\gamma$ are comparable, the Bayesian estimator needs more blocks to estimate the unknown phase. This is represented in Fig(2). We note that the feedback scheme could be implemented by employing fast control via field programmable gate arrays for CQED devices.
\begin{figure}
\begin{center}
\includegraphics[width=0.5\textwidth]{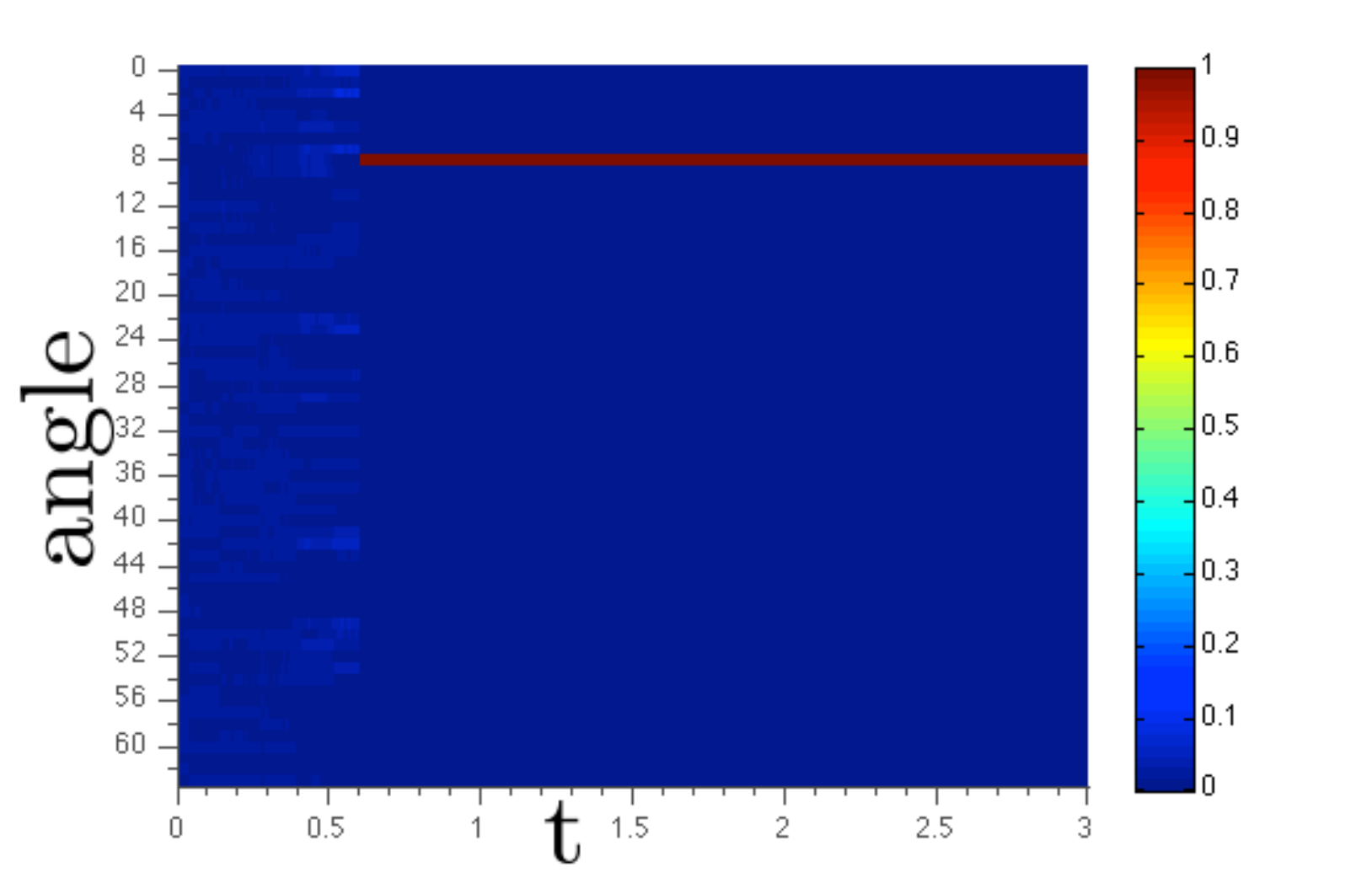}
\caption{\label{FIG1} Self-stabilizing phase measurements in the regime of good control: The qubit decoherence rates are two orders of magnitude smaller than the measurement strength. In this regime, if the qubit measurement were to have a preferred direction, the qubit would undergo zeno-like dynamics. In this regime, it is seen that the unknown phase is estimated correctly by the Bayesian estimator discussed in the text.}
\end{center}
\end{figure}
The essential part of this control scheme is the choice of approximately unbiassed measurements to rapidly purify a qubit in a rotating frame which is being estimated by a Bayesian estimator. From the standpoint of implementation, we can also introduce an additional delay between the end of the block and the implementation of the feedback loop. This would correspond to waiting for the feedback loop to be computed, a task that might not be slower than the ultrafast dynamics of qubit implementations such as CQED.
\begin{figure}
\begin{center}
\includegraphics[width=0.5\textwidth]{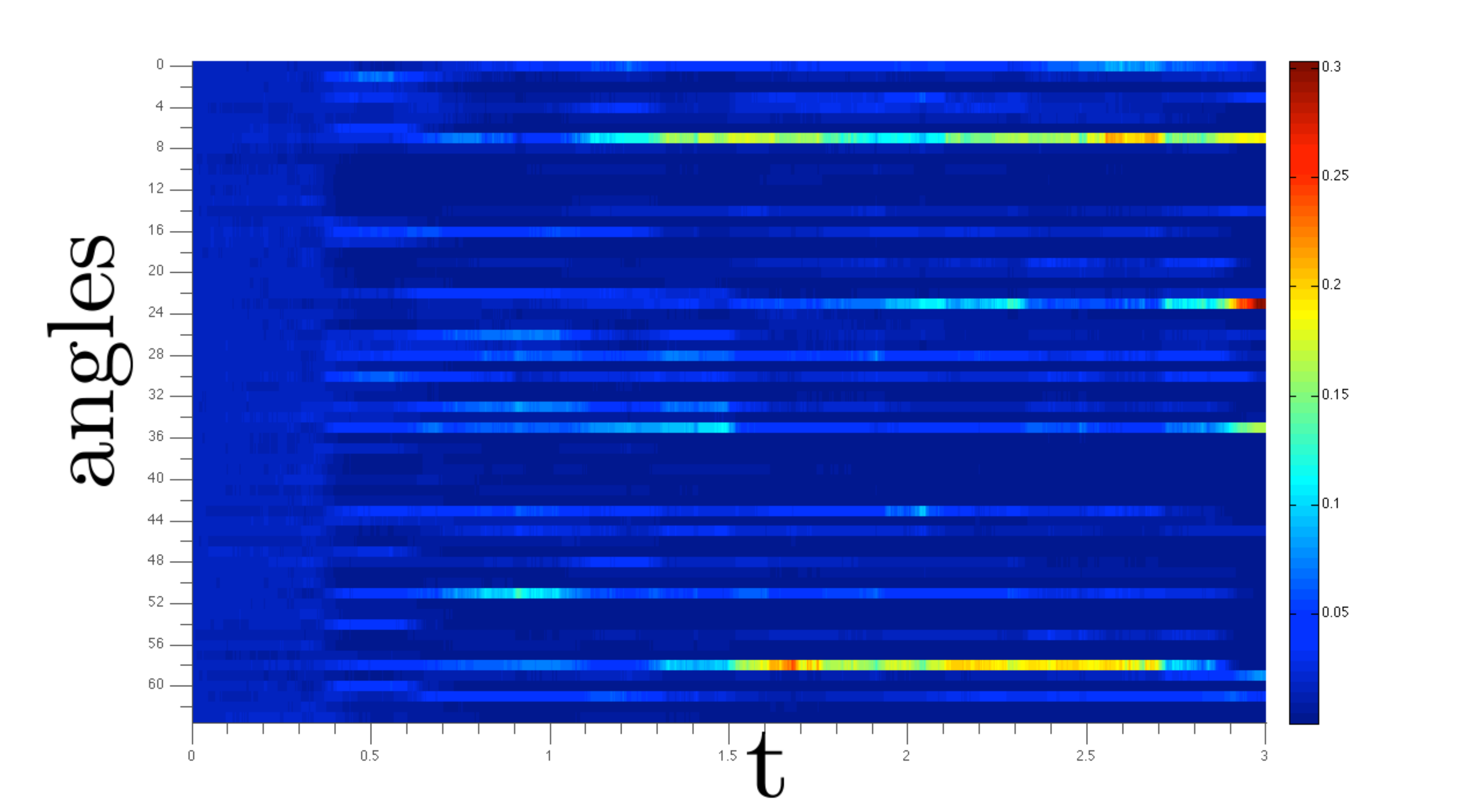}
\caption{\label{FIG1} Self-stabilizing phase measurements in the regime of bad control: The qubit decoherence rates are each comparable to measurement strength. In this regime, the qubit suffers strong decoherence and several cycles are needed to estimate the unknown phase.}
\end{center}
\end{figure}
In this article, we have demonstrated active feedback stabilization of a qubit decohering while acquiring an unknown phase. We employed continuous measurements to stabilize the qubit. By using Bayesian estimation in conjunction with rapid purification, we have simulated a qubit that is stabilized in the regime of good control. Finally, whether there exist local measurement and feedback schemes capable of self-stable metrology not with single qubit state, but entangled states, is an open question.

\begin{acknowledgments}
Centre for Quantum Technologies is a Research Centre of Excellence funded by the Ministry of Education and the National Research Foundation of Singapore.
\end{acknowledgments}
\bibliography{Vinjanampathy_04302014.bib}
\end{document}